\documentclass[12pt]{article}
\usepackage{graphicx}

\title{ Aspects of Relativistic Sum Rules}

\author{ Scott M. Cohen\\
Department of Physics, Duquesne University\\
Pittsburgh, PA  15282-0321\\
}
\begin{document}
\maketitle
\begin{abstract}

The status of our understanding of relativistic sum rules is reviewed. The recent development of new theoretical methods for the evaluation of these sum rules offers hope for further advances in this challenging field. These new techniques are described, along with a discussion of the source of difficulties inherent in such relativistic calculations. A connection is pointed out between certain sum rules for atomic interactions with charged particles and those for interactions with photons.

\end{abstract}

\newpage

\tableofcontents

\newpage

\section{Introduction}
\label{Intro}

In the theory of energy deposition from fast, charged particles and from photons, sums of quantum mechanical matrix elements weighted by a power of energy can play an important role \cite{InokutiRMP,Dalgarno,Jackiw}. While the matrix elements involved depend explicitly on the excited states of the system under consideration, these sums can often be shown to depend solely on properties of that system's ground state. Such a demonstration involves the use of closure, and the resulting expressions are known as \lq \lq sum rules".  In this article, I review work on these sum rules for the case of a relativistic target. 

As an illustration of such weighted sums of matrix elements, calculations of the average energy transfer in descriptions of scattering processes lead to sums of the form,

\begin{equation}
	\label{GenSum} 
	S = \sum_n E_{ni} \vert \langle n|A|i \rangle \vert^2,
\end{equation}

\noindent where the sum is over all eigenstates, $|n\rangle$ with energy $E_n$, of the Hamiltonian describing the target system, $A$ is an operator representing an interaction between incident particle and target, and $E_{ni} = E_n - E_i$, with $E_i$ the energy of the target's initial state $|i \rangle$. For a dipole interaction, $A = z$, Eq.~(\ref{GenSum}) becomes the well-known Thomas-Reiche-Kuhn (TRK) sum, whereas with $A = e^{i\vec q \cdot \vec r /\hbar}$ it is the Bethe sum, with $\vec q$ the momentum transferred to the target, and $z$ and $\vec r$ are position operators. A non-relativistic theory of photon absorption involves the interaction $A = \vec \epsilon \cdot \vec p e^{i \vec k \cdot \vec r}$, where $\vec \epsilon$ and $\hbar \vec k$ are the polarization and momentum of the absorbed photon, the magnitude of the latter being related to the energy transfer by $\hbar k= E_{ni} / c$. A relativistic theory for photoabsorption has a similar form, but the momentum operator $\vec p$ is replaced by the Dirac matrix, $\vec \alpha$.

In non-relativistic (NR) theories, evaluation of these sums is often quite simple. One need only remember the Schr\"odinger equation, $H |n\rangle = E_n |n \rangle$, and the closure relation, $ \sum_n | n \rangle \langle n | = {\cal I}$ with ${\cal I}$ the identity operator, where the sum extends over the complete set of states of the Hamiltonian. For a single particle bound in a static potential, we have the Schr\"odinger Hamiltonian,

\begin{equation}
	H_S = \frac{p^2}{2m} + V. 
\end{equation}

\noindent Then, using $A = z$, the TRK sum rule is

\begin{eqnarray}
	\label{TRK}
	S_{1} &  \equiv &\frac{2m}{\hbar^2} \sum_n E_{ni} \langle i|z| n \rangle \langle n|z|i \rangle \nonumber \\
			&  = &\frac{m}{\hbar^2} \sum_n \left(\langle i|z| n \rangle \langle n|[H_S,z]|i \rangle + \langle i|[z,H_S]| n \rangle \langle n|z|i \rangle \right) \nonumber \\
			& = & \frac{m}{\hbar^2} \langle i|[z,[H_S,z]]|i \rangle  =  1.
\end{eqnarray}

In this paper, I will discuss the effects of relativity on this and other sum rules. Unlike their NR counterparts, calculations of relativistic sum rules are fraught with difficulties. These difficulties can generally be traced to the presence of negative energy states, or from another point of view, the existence of positrons. Let us see what happens if we simply replace $H_S$  in Eq.~(\ref{TRK}) by the Dirac Hamiltonian,

\begin{equation}
	H_D = \left ( \beta mc^2 + c\vec \alpha \cdot \vec p + V \right ).
\end{equation}

\noindent Then,

\begin{eqnarray}
	\label{Dirac}
	S_{1}  = \frac{m}{\hbar^2} \langle i|[z,[H_D,z]]|i \rangle = 0,
\end{eqnarray}

\noindent a result of the fact that $H_D$ is linear in the momentum operator. There is obviously no way that this result can reduce to that obtained in the NR calculation, indicating that there is something seriously wrong. It is not difficult to show that transitions into final states $| n \rangle$ of negative energy have precisely canceled transitions into those with energies that are positive \cite{Levinger}. In a relativistic theory and assuming the system is initially in its ground state, the negative energy states are all filled (\lq \lq Dirac sea"), so that transitions into these states are specifically forbidden by Pauli's exclusion principle. Of course, our naive approach utilizing closure in the Dirac representation automatically includes these forbidden transitions. To obtain sensible results for relativistic sum rules, a method must be found to impose the exclusion principle by projecting out the negative energy states.

The case of photoabsorption has additional difficulties, since the operator $A$ appearing in the matrix elements then includes an exponential of $E_{ni}$. One might again try eliminating the explicit dependence on $E_n$ in favor of $H$, by

\begin{equation}
	e^{iE_{ni}z/\hbar c} | n \rangle = e^{i(H - E_{i})z/\hbar c} | n \rangle,
\end{equation}

\noindent but because of the fact that $H$ and $z$ do not commute, this relation is not even correct. Therefore, as will be discussed below, calculations of photoabsorption sum rules have resorted to expansions of the exponential in powers of $E_{ni} z$, under the assumption  of long-wavelengths, and keeping only the first few terms in the expansion.

The outline of the paper is as follows: In the next section, I discuss how studies of interactions between atoms and charged particles or photons lead to a consideration of sum rules. In Section~\ref{review}, I review early efforts to evaluate these sum rules using a relativistic model of the target. Following this in Section~\ref{recent}, I present the results of my own work on this subject, studying the relativistic Bethe sum rule in the context of stopping power or mean energy loss from charged projectiles. I also demonstrate a close relationship between other sums that arise in the context of charged particle scattering and those related to studies of photon absorption. In Section~\ref{relativity}, I consider relativistic sum rules from the perspective of Dirac's hole theory, and offer some thoughts on how the infinities present in this theory -- of energy and of the number of electrons -- lead to difficulties similar to those described above in the context of a single-particle theory. Finally, a conclusion is given in Section~\ref{remarks}.

\section{Origin of sum rules}
\label{sumrules}

In this section, I review basic ideas concerning interactions between atoms and either charged particles or photons, and show how sum rules arise in such contexts. I begin in Sec. \ref{particles} with a discussion of the interaction of fast, charged particles with matter, and then in Sec. \ref{photon} describe the case of photoabsorption.

\subsection{Scattering of fast, charged particles}
\label{particles}

In the first-Born approximation, the differential scattering cross-section (DCS) for an excitation of the target to a state of energy $E_n$ by a fast, charged projectile may be written in the Coulomb gauge as \cite{Fano}

\newpage

\begin{eqnarray}
	d\sigma_n & = & \frac{2\pi z^2 e^4}{mv^2} \left \{
		\frac{\left| F_n\right| ^2}{Q^2\left(1+Q/2mc^2\right)^2}  \right. \nonumber \\
		 && \left. + \frac{| \vec{\beta}_t \cdot \vec{G}_n | ^2}
			{\left[ Q \left( 1 + Q/2mc^2 \right) - E_{ni}^2/2mc^2 \right] ^2}
		\right \} \left(1 + \frac{Q}{mc^2} \right) dQ.
	\label{sigma}
\end{eqnarray}

\noindent Here, $\vec v$ is the initial velocity, and $z$ the atomic number, of the projectile, and $Q = \sqrt{m^2c^4 + q^2c^2} - mc^2$ is known as the recoil energy. The component of $\vec{\beta}=\vec v /c$ perpendicular to $\vec{q}$ is denoted by $\vec \beta_t$. The quantities, $F_n$ and $\vec{G}_n$ in Eq.~(\ref{sigma}), are defined by

\begin{eqnarray}
		\label{Fn}
	F_n & = & \sum_{j=1}^Z \langle n \vert e^{i \vec{q} 
				\cdot \vec{r}_j / \hbar} \vert i \rangle, \\
		\label{Gn}
	\vec{G}_n & = & \sum_{j=1}^Z \langle n \vert  
					e^{i\vec{q} \cdot \vec{r}_j / \hbar} \vec{\alpha}_j \vert i \rangle,
\end{eqnarray}

\noindent where $Z$ is the atomic number of the target atoms. The first term in Eq.~(\ref{sigma}) is referred to as the longitudinal interaction and may readily be derived using the static, unretarded Coulomb interaction. The other term, that containing $\vec{G}_n$, corresponds to the exchange of a single, transverse photon and is referred to as the transverse interaction.

Use of Eq.~(\ref{sigma}) leads to consideration of sums in various contexts. An important example is the calculation of the stopping power, or mean energy loss per unit pathlength from the projectile \cite{Bethe1,Bethe3}. This is found from $d\sigma_n$ as

\begin{equation}
	\label{stop}
	-\frac{dE}{dx} = {\sum_n}^{\, \prime} E_{ni} \int d\sigma_n,
\end{equation}

\noindent and the prime on the summation symbol means the sum is restricted by energy conservation, such that there is a maximum energy transfer, $E_{ni} < E_{Max}$. From Eq.~(\ref{sigma}), it is seen that the above equation involves an integral over the recoil energy. It turns out from kinematic considerations that the range of this integration is $Q_m < Q < Q_{Max}$, where the minimum is related to the energy transfer as $Q_m = E_{ni}^2/2mv^2$, and $Q_{Max} = 2mv^2(M/m)^2$, with $M$ the projectile mass \cite{InokutiRMP,Fano}. Since the sum does not include a complete set of states of the target system, closure may not immediately be used to perform the summation. Nonetheless, it has been argued \cite{Fano} that it is a good approximation to extend this sum to include all energies, $\sum^{\, \prime} \rightarrow \sum_n$, as long as one also replaces the upper limit of the $Q$-integral with $Q_M = 2mv^2/(1-v^2/c^2)$. The errors incurred by this replacement are part of the much studied shell corrections \cite{Walske1,Walske2,BichselShell,BetheRMP}.

When $v \ll c$, only the longitudinal interaction is important, and then

\begin{equation}
	-\frac{dE}{dx} = \sum_{n} E_{ni} \int_{Q_m}^{Q_M} \frac{\left| F_n(\vec q)\right| ^2}{Q^2\left(1+Q/2mc^2\right)^2} (1 + Q/mc^2) dQ.
\end{equation}

\noindent It is advantageous to perform the summation before the integration, but one must first account for the fact that the lower limit of integration depends explicitly on $E_n$. A common approach is to split the $Q$-integral into two parts: the low- and high-$Q$ regimes, separated by a cutoff, $Q_c$. Then the integration limits for the high-$Q$ regime do not depend on the energy of the excited state, and the order of summation and integration may be interchanged in the expressions for these integrated quantities. Thus, we are led to consideration of the Bethe sum \cite{BetheJackiw},

\begin{equation}
	S_{1}(Q) = \sum_{n} E_{ni} |F_n(\vec q)|^2 = \sum_{j=1}^Z \sum_{n} E_{ni} \left| \langle n \vert e^{i \vec{q} \cdot \vec r_j / \hbar} \vert i \rangle \right|^2,
 	\label{S1Q}
\end{equation}

\noindent where I have followed Bethe in neglecting the small correlation terms involving two different electrons.

For the low-$Q$ region, the integration must be done first. To accomplish this, a long-wavelength approximation is introduced by expanding the exponential in $F_n(\vec q)$ as $e^{i\vec q\cdot \vec r_j} \approx 1 + i\vec q\cdot \vec r_j$. Then, neglecting terms ${\cal O}(Q/mc^2)$ gives $F_n(\vec q) \approx Qf_n/E_{ni}$, with the optical dipole oscillator strength defined as

\begin{equation}
 	\label{fn}
	f_n = \frac{2mE_{ni}}{\hbar^2} \left|\sum_j \left\langle n\left|z_j\right|i \right\rangle\right|^2.
\end{equation}

\noindent This leads to the TRK sum rule, which for a system of $Z$ electrons is

\begin{equation}
 	\label{S1}
	S_1 = \sum_{n} f_n = \lim_{q \rightarrow 0} \frac{2m}{q^2} S_1(Q) = Z.
\end{equation}

For a relativistic projectile, the transverse part of the DCS must also be included. The appearance of $E_{ni}$ in the denominator of this term complicates its treatment in terms of sum rules. I have recently shown how to treat the high-$Q$ case by expanding the denominator in powers of the small quantity $(Q - E_{ni})/Q$ \cite{stop_one}, leading to sums of the form

\begin{equation}
	\label{tsums}
	{\cal T}^{(\nu)} (Q) =  \sum_{j=1}^Z \sum_{n} (Q - E_{ni})^\nu \left| \langle n |e^{i \vec q \cdot \vec r_j / \hbar} \vec \beta_t \cdot \vec \alpha_j | i \rangle \right|^2.
\end{equation}

For the low-$Q$ case, expansion of the denominator is not appropriate, since $Q -  E_{ni}$ and  $Q$ may be comparable in magnitude. When the target system is non-relativistic, Fano showed that the transverse contribution is proportional to $S_1$ for small $Q$. In the case of a relativistic target, I have also shown \cite{stop_one} that Fano's approach must be generalized, as will be discussed in Section~\ref{trans}. 

\subsection{Photon interactions}
\label{photon}

The study of sum rules for photon-matter interactions grew out of a conjecture by Gell-Mann, Goldberger, and Thirring (GGT) \cite{GGT}. In the course of investigating the implications of causality on quantum theory and, in particular, on scattering amplitudes derived from field theory, these authors arrived at a dispersion relation between the dispersive and absorptive parts of the photon forward scattering amplitude. Let us now look at how this leads to sums that are quite similar to those discussed in the previous section.

In their paper, GGT considered the matrix element of the $S$ matrix, $S = 1 - iR$, for scattering of a photon of four-momentum $k^{\prime \prime}$ and polarization $\sigma$ into one of four-momentum $k^\prime$ and polarization $\lambda$, by a matter system which is ultimately left with its energy unchanged. They wrote,

\begin{eqnarray}
	\label{R}
	\langle \lambda k^\prime f|R|i \sigma k^{\prime \prime} \rangle & = & -i \int_{-\infty}^\infty d^4x^\prime \int_{-\infty}^\infty d^4x^{\prime \prime } \langle \lambda k^\prime f|V(x^\prime) V(x^{\prime \prime})|i \sigma k^{\prime \prime} \rangle \eta(t^\prime-t^{\prime \prime})  \nonumber \\
		 & \equiv & F_{\lambda\sigma}^{fi}(k^\prime,k^{\prime \prime}) ,
\end{eqnarray}

\noindent with the step function $\eta(t^\prime-t^{\prime \prime}) $ vanishing unless $t^\prime > t^{\prime \prime}$, in which case it is unity. The interaction potential is

\begin{equation}
	V(x) = -\vec j(x) \cdot \vec A(x),
\end{equation}

\noindent where the current operator $\vec j$ and the transverse vector potential,

\begin{equation}
	\vec A(x) = \int d^3 k \sum_{s=1}^2 \left[a_{\vec k s}\vec \epsilon_s e^{i(\vec k \cdot \vec x- \omega t)} +a_{\vec k s}^\dagger\vec \epsilon_s e^{-i(\vec k \cdot \vec x- \omega t)} \right],
\end{equation}

\noindent are operators in the interaction representation. The operator $a_{\vec k s}^\dagger (a_{\vec k s})$ creates (annihilates) a photon with momentum $\hbar \vec k$, frequency $\omega = kc$, and polarization $\vec \epsilon_s$ perpendicular to $\vec k$.

Since the initial and final states each have a single photon present and $V$ creates or annihilates one photon, it is clear that in Eq.~(\ref{R}), there are two possible ways for the system to pass from the initial to the final state. Either a photon is first emitted and then one is later absorbed, or vice-versa. This yields two separate terms  in Eq.~(\ref{R}), from which the scattering amplitude becomes \cite{GGT}

\begin{eqnarray}
	\label{F}
	F_{\lambda\sigma}^{fi}(k^\prime,k)  & =  & \sum_n \int d^3x^\prime \int d^3x^{\prime\prime} \left[\frac{\langle f|j_\lambda   (\vec x^{\, \prime})| n \rangle \langle n | j_\sigma (\vec x^{\, \prime\prime})|i \rangle }{E_i + \hbar \omega-E_n+i\epsilon}e^{-i\vec k \cdot \vec x^\prime +i \vec k^\prime \cdot \vec x^{\prime \prime}} \right. \nonumber \\
		&& + \left.\frac{\langle f|j_\sigma  (\vec x^{\, \prime})| n \rangle \langle n | j_\lambda (\vec x^{\, \prime\prime})|i \rangle }{E_i - \hbar \omega-E_n+i\epsilon}e^{i\vec k \cdot \vec x^\prime -i \vec k^\prime \cdot \vec x^{\prime \prime}}\right].
\end{eqnarray}

By considering the analytic properties of this quantity, GGT derived the following dispersion relation between the real (${\cal R}e$) or dispersive and the imaginary (${\cal I}m$) or absorptive parts of the forward scattering amplitude for photons of frequency $\omega$:

\begin{eqnarray}
	\label{dispersion}
	{\cal R}e f(\omega) - {\cal R}e f(0) &  = & \frac{2\omega^2}{\pi} P \int_0^\infty d\omega^{\, \prime} \frac{{\cal I}m f(\omega^{\, \prime})}{\omega^{\, \prime} ({\omega^{\, \prime}}^2 - \omega^2)} \nonumber \\
			& = &  \frac{\omega^2}{2\pi^2} P \int_0^\infty d\omega^{\, \prime} \frac{\sigma_{abs}(\omega^{\, \prime})}{{\omega^{\, \prime}}^2 - \omega^2},
\end{eqnarray}

\noindent where $f(\omega) = F_{\lambda \lambda}^{ii}(k,k)$, $P$ denotes the principal value of the integral, and use has been made of the optical theorem, ${\cal I}m f(\omega) =\omega \sigma_{abs}(\omega)/4\pi$, with $\sigma_{abs}(\omega)$ the total absorption cross section for a photon of frequency $\omega$. For scattering from bound electrons, ${\cal R}e f(0) = 0$ (Rayleigh scattering). Then, with the conjecture that for $\omega \rightarrow \infty$, $f(\omega)$ approaches its free electron value,

\begin{equation}
	\label{conjecture}
	{\cal R}e f(\infty) = -e^2/m, 
\end{equation}

\noindent GGT obtain the sum rule,

\begin{equation}
	\label{GGTsum}
	\int_0^\infty d\omega \, \sigma_{abs}(\omega) = \frac{2\pi^2 e^2}{m}.
\end{equation}

As I will now demonstrate, this is closely related to the TRK sum rule, Eq.~(\ref{TRK}). For a single non-relativistic electron, we have 

\begin{equation}
	\vec j(\vec x) = \frac{e}{2m} \{\vec p ,\delta(\vec x - \vec r)\}
\end{equation}

\noindent Then the total photoabsorption cross section may be found directly from Eq.~(\ref{F}) and the optical theorem, and is given by

\begin{eqnarray}
	\label{abs}
	\sigma_{abs}(\omega) = \frac{4 \pi^2 e^2}{m^2\omega} \sum_n | \langle n | p_{x} e^{ikz} | i \rangle |^2 						[\delta(\hbar\omega - E_{ni}) + \delta(\hbar\omega + E_{ni})],
\end{eqnarray}

\noindent where I have taken the polarization $\lambda$ along the $x$ direction and the momentum $\vec k$ along $z$. If the system is initially in its ground state, $E_{ni} > 0$, and the second delta function will not contribute for positive frequencies. Then, expanding the exponential, we obtain

\begin{eqnarray}
	\label{pz}
	\int_0^\infty d\omega \, \sigma_{abs}(\omega) = \frac{4 \pi^2 e^2\hbar}{m^2} \sum_n \frac{1}{E_{ni}} \left (\left \vert\langle n | p_{x} | i \rangle \right \vert^2 +\frac{E_{ni}^2}{\hbar^2c^2}\left \vert\langle n | p_{x} z | i \rangle \right \vert^2\right. \nonumber \\
		 \left. - \frac{E_{ni}^2}{2\hbar^2c^2}[\langle n | p_{x} | i \rangle \langle n | p_{x} z^2 | i \rangle + H.c. ] \right ),
\end{eqnarray}

\noindent with $H.c.$ the hermitian conjugate. The first term in parentheses is the dipole term $E1$, the next is the quadrupole $E2$, and the remaining terms represent retardation effects on the dipole interaction, denoted as $E1^{ret}$.

In the dipole approximation only the first term is kept, and using the identity $p_{x} = im [H,x] / \hbar$, we see that the GGT sum rule, Eq.~(\ref{GGTsum}), becomes

\begin{equation}
	\frac{2m}{\hbar^2}\sum_n E_{ni} |\langle n | x | i \rangle |^2 = 1,
\end{equation}

\noindent which is just the dipole oscillator strength sum rule of TRK, as claimed. The important point is that  Eq.~(\ref{GGTsum}) is much more general in that it has been derived without the assumptions of long wavelengths, a dipole interaction, and a non-relativistic matter system. Unfortunately, as we will discuss in the following section, the conjecture Eq.~(\ref{conjecture}) is incorrect and so therefore is the GGT sum rule.

\section{Review of early work}
\label{review}

The conjecture of GGT was based on the notion that for very high energy photons, one might reasonably suppose that the binding energy of the electron will have little import, and can therefore be neglected. Nonetheless, GGT expressed concern about the correctness of this conjecture. In particular, they pointed out that the photoelectric cross section was then believed to have a $1/\omega$ dependence at high frequencies, leading to a logarithmic $\omega$-dependence on the right-hand side of Eq.~(\ref{dispersion}).

The apparent divergence of ${\cal R}ef(\omega)$ at large $\omega$ can be circumvented by introducing a procedure that was originally suggested by Brown \cite{BrownPrComm}. This procedure involves defining the relevant cross section as the difference between the photoelectric cross section and $\sigma_{pps}$, the cross section for pair production in which the electron of the pair is produced in the occupied $1s$ state. As was later discussed in detail by Erber \cite{Erber}, this definition of the cross section corresponds to scattering by the bound electron, alone. That is, it is in fact the difference between the cross section for a one-electron atom and that for an isolated proton \cite{Erber}. Erber than argues that the logarithmic dependence of the dispersive amplitude that was mentioned above is exactly canceled by this subtraction procedure. It should be noted that along with this subtraction of cross sections, one must include a corresponding subtraction of forward scattering amplitudes, which has also been carefully discussed by Erber \cite{Erber}. As is shown in his Eq.~(3.1), the correct amplitude is not then obtained directly from the scattering amplitude of GGT, our Eq.~(\ref{F}). Rather, by carefully considering the various processes involved in scattering from the atom and also from the free proton, and in particular, the processes involving creation and annihilation of electron-positron pairs, Erber has argued that the correct amplitude requires the replacement of $i\epsilon$ by $-i\epsilon$ in the second term of our Eq.~(\ref{F}) (compare Erber's Eq.~(3.1)).

Shortly after publication of the paper by GGT, Levinger and co-workers, using the Brown/Erber procedure, showed that the conjecture, Eq.~(\ref{conjecture}), still turns out to be incorrect \cite{Levinger,LevPayne,LevRustgi}. These authors began \cite{LevPayne,LevRustgi} with numerical studies of the dipole oscillator strengths for a single electron in the field of a $Pb$ nucleus undergoing transitions from the $1s$ into higher discrete states. They found that the retarded, relativistic oscillator strengths for these discrete transitions were only about $80 \%$ of their non-relativistic values \cite{LevPayne}. For transitions to the continuum states, they obtained values from the work of Brown \cite{Brown1,Brown2} and Hulme \cite{Hulme}. Including Jacobsohn's non-relativistic calculations of the electric quadrupole transitions ($E2$) \cite{Jacobsohn}, they found a summed oscillator strength of approximately $0.86$, in stark contrast to the value of unity predicted by GGT (inclusion of retardation in $E2$ changes this value to $0.87$ \cite{Levinger}). They argued that this result should be accurate to within a few percent, even considering the neglect of higher multipoles.

The following year, Levinger, Rustgi, and Okamoto addressed this question through the evaluation of sum rules \cite{Levinger}. To overcome the difficulties arising from the existence of negative energy states, these authors used a semi-relativistic Hamiltonian, which does not have the negative energy eigenstates of the original Dirac Hamiltonian. The important difference between their Hamiltonian and $H_S$ is the inclusion of the mass-variation term, $H_m = -p^4/8m^3c^2$ (other correction terms in their Hamiltonian are at most linear in the momentum operator, and therefore do not contribute to the sum rule).  Since the negative energy states have already been removed, the Brown/Erber subtraction procedure discussed above is not appropriate here (nor is it needed in the related sum rule calculations discussed below), since that procedure is specifically related to pair production, which is not included in this approach.

For the unretarded $E1$ term, $H_m$ adds a relativistic correction of (see Eq.~(\ref{TRK}))
 
\begin{equation}
	E1^{rel} = -\frac{1}{8m^2c^2\hbar^2} \langle i|[z,[p^4,z]]|i \rangle = -\frac {5}{3mc^2}T_0
\end{equation}

\noindent for an isotropic initial state, with $T_0 = \langle i | p^2/2m | i \rangle$. Using this same Hamiltonian, the $E2$ contribution was calculated relativistically, as were the retardation effects, $E1^{ret}$, on the dipole transitions. They found

\begin{equation}
	E2 = \frac {2}{3mc^2}T_0  + \frac{1}{15mc^2} \left \langle i \left|\left(r^2 \frac{d^2V}{dr^2} + 4 r \frac{dV}{dr}\right) \right| i \right\rangle,
\end{equation}

\noindent and

\begin{equation}
	E1^{ret} = -\frac{1}{15mc^2} \left\langle i \left|\left(r^2 \frac{d^2V}{dr^2} + 4 r \frac{dV}{dr}\right)  \right| i \right\rangle.
\end{equation}

\noindent The terms involving $V$ in $E2$ are exactly canceled by $E1^{ret}$. This  is one illustration of the well-known fact that higher multipoles and retardation each contribute at the same order of magnitude as relativistic effects, and that it is inconsistent to include one without the others \cite{Pratt1,Pratt2}.

Combining results, these authors obtained

\begin{equation}
	\label{LevSum}
	f(\infty) = -\frac{e^2}{m} \left(1 - \frac{T_0}{mc^2}\right) \approx - \frac{e^2}{\left(m + T_0/c^2\right)}
\end{equation}

\noindent for the GGT sum rule. They pointed out that the presence of $T_0$ in the denominator represents the increase of mass due to the electron's kinetic energy, consistent with the fact that its presence results directly from inclusion of $H_m$ in the Hamiltonian. To obtain a numerical result, they used the value, $T_0/mc^2 = Z^2\alpha^2/2 = 0.18$ for $Pb$, with $\alpha$ the fine structure constant. This yields a summed oscillator strength of $0.82$, in satisfactory agreement with their earlier result of $0.87$ \cite{LevRustgi}. They observed that exact agreement should not be expected, due to the approximate nature of their numerical calculations \cite{LevPayne,LevRustgi}, as well as the neglect of contributions ${\cal O}(Z^4\alpha^4)$ in the sum rule calculations.

The above result was confirmed a decade later by Goldberger and Low \cite{GoldLow}, who calculated the forward scattering amplitude in the limit of very high photon energies. They obtained the result,

\begin{equation}
	\label{GoldLowsum}
	f(\infty) = - \frac{e^2}{m} (1 - \frac{Z^2\alpha^2}{2} + \frac{5Z^4\alpha^4}{24} + \cdots),
\end{equation}

 \noindent which is in agreement with Eq.~(\ref{LevSum}) to lowest order. An exact expression for $f(\infty)$ in terms of quadratures was also derived \cite{GoldLow}.

Over the next several years, a number of papers considered the integrated photoabsorption cross section \cite{Dogliani,Friar,Matsuura,Gerasimov,Drake}, using analytical methods based on closure and generalizing the sum rules by: (a) including a world-scalar potential and utilizing the Foldy-Wouthuysen transformation (FWT) \cite{Foldy} to obtain a semi-relativistic Hamiltonian \cite{Friar}, and (b) studying a many-electron system with interactions \cite{Matsuura}. In spite of Levinger's work \cite{Levinger,LevPayne,LevRustgi}, Gerasimov concluded that the GGT sum rule, Eq.~(\ref{GGTsum}), was in fact correct \cite{Gerasimov}. By studying specific models, Matsuura \cite{Matsuura} obtained results in agreement with Levinger's \cite{Levinger}, and pointed out that Gerasimov had incorrectly assumed that $f(\omega)$ is analytic everywhere except on the real $\omega$ axis, invalidating his conclusions.

In 1985, Schmitt and Arenh\"ovel \cite{Schmitt} studied the GGT sum, using both analytical and numerical techniques. For the analytical efforts, they studied the leading relativistic corrections to the sum rule in three different models of the target particle: (a) non-relativistic (NR) or Schr\"odinger, (b) Foldy-Wouthuysen (FW)  using the FWT \cite{Foldy} to obtain a semi-relativistic model, and (c) Dirac using $H_D$ and introducing projection operators to exclude contributions from the negative energy states. In the latter case, they used an expansion of the projection operators in powers of $H_D / mc^2$, so those results are also semi-relativistic. Their use of the FWT to transform the multipole operators for the FW model represented a technical improvement over previous efforts \cite{Levinger,Friar}, wherein only the Hamiltonian and states had been transformed. Nonetheless, all their results were in agreement with that of the earlier work.

These authors also numerically calculated $f(\infty)$, using the exact (fully relativistic) expression of Goldberger and Low (see Eq.~(4.1) of \cite{GoldLow}) for a particle in a square well. Their results were such that $f(\infty)$ and the FW calculation of the GGT sum differed by an insignificant amount (less than $0.1 \%$). The implications of this result for our understanding of relativistic effects is not completely clear, however, since their chosen binding energies are rather small, being at most about $5 \%$ of the particle's rest energy. Nonetheless, they did find a small, but significant, contribution to the GGT sum from the pair production term, $\sigma_{pps}$. 

Shortly thereafter, Leung, Rustgi, and Long \cite{Leung+} considered relativistic corrections to the Bethe sum rule. In a non-relativistic theory, this sum can be readily evaluated exactly including all multipoles:

\begin{eqnarray}
		S_{1}(Q)  = \sum_{n} E_{ni} \left| \langle n \vert e^{i \vec{q} \cdot \vec r / \hbar} \vert i \rangle \right|^2  = \langle i \vert e^{-i \vec{q} \cdot \vec r / \hbar}  [H_S, e^{i \vec{q} \cdot \vec r / \hbar}] \vert i \rangle   =  Q,
 	\label{BSR}
\end{eqnarray} 

\noindent where I have again assumed an isotropic initial state. To obtain the relativistic corrections, Leung, et. al. \cite{Leung+} used an approach similar to that of \cite{Friar}, performing the FWT on $H_D$ for a one-electron system, obtaining

\begin{eqnarray}
	\label{LeungSum}
	S_1(Q) = Q \left ( 1 - \frac{5}{6m^2c^2} \langle \widetilde i \vert p^2 \vert \widetilde i \rangle \right ) ,
\end{eqnarray}

 \noindent  where $| \widetilde i \rangle$ is the transformed initial state, and terms ${\cal O}(Q^3)$ and ${\cal O}(Q^2p^2)$ were neglected. Leung later used this result \cite{Leung} to obtain the leading relativistic corrections to the Bethe stopping power. Note that this result is obtained by transforming the Hamiltonian and states, alone, while leaving the operator $e^{i \vec q \cdot \vec r / \hbar}$ unchanged. Thus, this procedure is inconsistent in its treatment of the various operators. We can also see that the result is incorrect by comparing to that for a relativistic free-particle at rest \cite{Fano},

\begin{equation}
	\label{free}
	S_1^0(Q) = \frac{Q(1 + Q/2mc^2)}{(1 + Q/mc^2)} \approx Q(1 - Q/2mc^2).
\end{equation}

\noindent The result of Eq.~(\ref{LeungSum}) should reduce to this expression if the expectation value of $p^2$ is set to zero, and is thus seen to be missing the term proportional to $Q^2$. A proper treatment using the FWT and transforming all operators was later formulated \cite{mysumrule}, obtaining

\begin{equation}
	\label{lowQ}
	S_1(Q) = Q \left ( 1 - \frac{5}{6m^2c^2} \langle \widetilde i \vert p^2 \vert \widetilde i \rangle - Q^2/2mc^2\right ),
\end{equation}

\noindent in agreement with Eq.~(\ref{free}). Other sums, involving various moments of the energy transfer, were also evaluated \cite{mysumrule}.

It was suggested in \cite{Leung+} that use of a no-pair Hamiltonian \cite{Sucher} and/or field-theoretic techniques might be useful for consideration of many-electron atoms. Such techniques were later used by Aucar, Oddershede, and Sabin \cite{Sabin} in their work on the relativistic corrections to the Bethe sum. They introduced a second-quantized no-pair Hamiltonian and projected the operator $e^{i \vec q \cdot \vec r_j/\hbar}$ onto the space of positive energy states, thereby excluding transitions into the negative energy states. These authors mention that it is straightforward to use their approach for the case of a many-electron atom, but they explicitly consider only the one-electron case. Unfortunately, although the techniques used in this approach are formidable, it does not appear to the present author that there is any simple way to obtain detailed results from their expressions, which are still in the form of a sum over a complete set of eigenstates of the Hamiltonian.

\section{Recent advances}
\label{recent}

\subsection{Relativistic Bethe sum rule}

In this section, I describe the work I have recently done on the relativistic Bethe sum rule. I will discuss the FWT \cite{Foldy}, derive a formally exact expression for the sum rule, and then explain why the results of \cite{Leung+,mysumrule} are only valid when $q$ is small. I will also show how to obtain an exact treatment of the $q$-dependence of the Bethe sum rule, using an FWT-type transformation introduced by Douglas and Kroll \cite{DouglasKroll}. When combined with the evaluation of sum rules for the transverse part of the DCS, these results permit the calculation of relativistic corrections to the stopping power that are valid for a fully relativistic projectile \cite{stop_one}. At the end of this section, I outline a method that appears to give highly accurate expressions for the TRK and Bethe sums even for very strongly bound systems.

To implement the FWT, one inserts the identity operator in the form of $e^{-iU}e^{iU}$, with $U$ chosen to be Hermitian so that $e^{iU}$ is unitary. For this preliminary discussion, I will assume that $U$ is the exact FWT operator for a general many-electron Hamiltonian $H$. Then, for example,

\begin{eqnarray}
	\langle n|He^{i \vec{q} \cdot \vec r_j / \hbar} |i \rangle =  \langle n|e^{-iU}e^{iU} He^{-iU}e^{iU} e^{i \vec{q} \cdot \vec r_j / \hbar} e^{-iU}e^{iU} |i \rangle.
\end{eqnarray} 

\noindent Thus, $e^{iU}$ induces a canonical transformation of the states and operators, with the identification, $\vert \widetilde n  \rangle = e^{i U} \vert n \rangle$ and $\widetilde H  = e^{i U} H e^{-i U}$. Then, defining $\vec q$-dependent operators through the replacement of $\vec p_j$ by $\vec p_j + \vec q$ due to a transformation of the form, ${\cal M} \rightarrow {\cal M}(\vec q) = e^{- i \vec{q} \cdot \vec{r_j} / \hbar} {\cal M} e^{i \vec{q} \cdot \vec{r_j} / \hbar}$, we have 

\begin{eqnarray}
	\langle n|He^{i \vec{q} \cdot \vec r_j / \hbar} |i \rangle  = \langle \widetilde n|e^{i \vec{q} \cdot \vec r_j / \hbar} \widetilde H(\vec q)\left(e^{iU(\vec q)} e^{-iU} \right)|\widetilde i \rangle.
\end{eqnarray} 

\noindent The FWT operator, $e^{i U}$, is chosen so as to bring the Hamiltonian into a \lq \lq block diagonal" form, explicitly separating the positive and negative energy subspaces. Now these positive energy states are still $4$-component spinors, but their third and fourth components vanish. Therefore, though $e^{iU(\vec q)} e^{-iU}$ is a $4 \times 4$ operator, only its upper-left $2 \times 2$ block will contribute to the matrix elements in our sum. Defining this block to be $R_j$, we then have,

\begin{eqnarray}
	\langle n|[H,e^{i \vec{q} \cdot \vec r_j / \hbar}] |i \rangle  = \langle \widetilde n|e^{i \vec{q} \cdot \vec r_j / \hbar} (\widetilde H(\vec q)R_j - R_j H)|\widetilde i \rangle,
\end{eqnarray} 

\noindent and

\begin{eqnarray}
	\langle i|e^{-i \vec{q} \cdot \vec r_j / \hbar} |n \rangle  = \langle \widetilde i|R_j^\dagger e^{-i \vec{q} \cdot \vec r_j / \hbar}|\widetilde n \rangle,
\end{eqnarray} 

\noindent which when combined and summed, yield a formally exact expression \cite{stop_one} for the Bethe sum, Eq.~(\ref{S1Q}), 

\begin{equation}
		\label{myBethesum}
		S_1(Q) = \sum_{j=1}^Z \langle \widetilde i \vert  R_j^ \dagger \left(\widetilde H (\vec q) R_j
		 - R_j \widetilde H \right)  \vert \widetilde i \rangle.
\end{equation}

To this point, I have assumed that $U$ is the exact FWT operator. However, given that this operator is not known for the case of bound electrons, it will eventually be necessary to introduce approximations for $e^{iU}$. Nonetheless, the fact that Eq.~(\ref{myBethesum}) is an exact result for the sum rule implies that the excited states have been treated exactly, and an approximate treatment of the operators appearing in the resultant expectation value does not alter this conclusion. In particular, this expression remains valid even when a semi-relativistic approximation to the FWT is used under circumstances for which the original summation includes significant contributions from highly relativistic states $|n \rangle$, such as with very large values of $Q$.

Let us now see why the approximations used in \cite{Leung+,mysumrule} for a single electron yield results that are only valid when $q$ is small. These authors used the approximation of \cite{Foldy},

\begin{equation}
	\label{FWexp}
	e^{i U} = \cdots e^{i U_3} e^{i U_2} e^{i U_1},
\end{equation}

\noindent which involves an expansion in powers of the small quantities $p /mc$ and $V /mc^2$, with $U_{N} \sim {\cal O}(1/m^{N})$. The calculation of the FW Hamiltonian \cite{Foldy} then proceeds by an expansion in powers of the $U_N$. For example, the first step of the FWT gives

\begin{equation}
	\label{H1}
	H_1 \equiv e^{i U_1} H e^{-i U_1} = H +i[U_1,H] - \frac{1}{2} [U_1,[U_1,H]] + \cdots,
\end{equation}

\noindent which is an expansion in powers of $p/mc$, since $U_1 = -i\beta \vec \alpha \cdot \vec p / 2mc$. Higher order corrections to the Hamiltonian are obtained from $U_2$ and $U_3$ in a similar fashion \cite{Foldy}. Now, $H_1(\vec q)$ is obtained from $H_1$ by the replacement of $\vec p$ by $\vec p + \vec q$. This means that the above expansion in powers of $p/mc$ leads directly to an expansion in powers of $q/mc$. Since this approach has started out with a $p$-expansion, it ends up with a $q$-expansion and is therefore incapable of handling large values of $q$. Therefore, unlike the NR Bethe sum rule, the relativistic corrections of \cite{Leung+,mysumrule} do not include contributions from all multipoles. In fact, the difference between the free and bound particle results, given in Eqs.~(\ref{free}) and (\ref{lowQ}), arises only from the dipole contributions. Therefore, the latter equation gives us little more information than the corrections to the TRK sum rule.

An expansion in powers of the potential energy $V$ does not lead to the difficulty described in the previous paragraph, at least as long as $V$ commutes with $e^{i\vec q \cdot \vec r / \hbar}$ so that $e^{-i\vec q \cdot \vec r / \hbar}Ve^{i\vec q \cdot \vec r / \hbar} = V$. This indicates that the way to circumvent the restriction to low $q$ is to avoid, or at least delay, the expansion in powers of $p$. Douglas and Kroll \cite{DouglasKroll}, have introduced a modified FWT that only expands in powers of $V$ \cite{DouglasKroll,Hess1,Hess2} and is therefore suitable for this purpose. This transformation (DKT), which is still of the form of Eq.~(\ref{FWexp}), takes $U_1 = U_f$ for a many-electron system, where

\begin{equation}
	\label{FWT}
	e^{i U_f} = \prod_{k=1}^Z e^{i U_k},
\end{equation}

\noindent and the free-particle FW operator for the $k^{th}$ electron is given by

\begin{eqnarray}
		\label{FWfree}
	e^{\pm i U_k} = \frac{1}{\sqrt{2\epsilon_k(\epsilon_k + mc^2)}}(\epsilon_k + mc^2 \pm  \beta_k \vec \alpha_k \cdot \vec p_k),
\end{eqnarray}

\noindent with $\epsilon_{k} = \sqrt{p_k^2c^2 + m^2c^4}$. 

For $H$, I have used \cite{stop_one} a no-pair many-electron Hamiltonian \cite{Sucher},

\begin{equation}
	H = \sum_{k = 1}^Z H_D^k +  {\cal L}_{+} \left( \sum_{k > l}^Z V_{kl} \right) {\cal L}_{+},
\end{equation}

\noindent where $H_D^k$ is the single-electron Dirac Hamiltonian for electron $k$, $V_{kl}$ is the interaction (Coulomb-Breit, for example) between electrons $k$ and $l$, and ${\cal L}_{+} = \prod_{k=1}^Z {\cal L}_{+}(k)$ with ${\cal L}_{+}(k)$ the projection operator onto the space of positive energy states of $H_D^k$.  It is true that the $V_{kl}$ part of the potential energy does not commute with $e^{i\vec q \cdot \vec r_j /\hbar}$, so it is not precisely correct to claim that a $V$-expansion will yield the exact $q$-dependence for this Hamiltonian. Nevertheless, the contributions to the sum rule from $V_{kl}$ (apart from its effect on the eigenstate $|\widetilde i \rangle$, which will not depend on $q$) turn out to be quite small \cite{stop_one,myRange}, so the error incurred by omitting this $q$-dependence should be negligible.

The first step of the DKT transforms $H$ into

\begin{equation}
	\label{H}
	\widetilde H = \sum_{k=1}^Z \epsilon_{k} + \widetilde V.
\end{equation}

\noindent where I have written $\widetilde V = e^{iU_f}(\sum_{k=1}^Z V_k + {\cal L}_{+} \sum_{k > l}^Z V_{kl} {\cal L}_{+})e^{-iU_f}$ to represent the potential energy terms. As I will discuss below, these terms appear to lead to very small corrections, so will not be further considered in any explicit manner. Using this as our transformed Hamiltonian, we have

\begin{equation}
	\label{Hq}
	\widetilde H(\vec q) = \sum_{k \ne j}^Z \epsilon_{k} + \epsilon_j(\vec q) + \widetilde V.
\end{equation}

\noindent Furthermore,

\begin{equation}
	e^{i U_f(\vec q)} = e^{iU_j(\vec q)} \prod_{k \ne j}^Z e^{i U_k},
\end{equation}

\noindent and

\begin{equation}
	\label{Ujq}
	e^{iU_j(\vec q)} = \frac{1}{\sqrt{2\epsilon_j(\vec q)(\epsilon_j(\vec q) + mc^2)}}(\epsilon_j(\vec q) + mc^2 \pm  \beta_j \vec \alpha_j \cdot (\vec p_j + \vec q)).
\end{equation}

All these expressions may now be incorporated into Eq.~(\ref{myBethesum}) and expanded in powers of $p_j$, while keeping the exact dependence on $q$. When this is done, one finds the relativistic Bethe sum rule to be \cite{myRange}

\begin{equation}
	\label{S1q}
	S_1(Q) = ZS_1^0(Q)( 1 - \Delta_f(Q) - \Delta_{_V}(Q)),
\end{equation}

\noindent with

\begin{eqnarray}
	\label{deltaf}
	\Delta_f(Q)  & =  & \frac{1}{3m^2c^2 Z \left( 1 + Q/mc^2 \right)^2}\left(1 + \frac{3}{2(1 + Q/mc^2)^2} \right) \sum_{j = 1}^Z \langle \widetilde 0 \vert p_j^2 \vert \widetilde 0 \rangle \nonumber \\
			&&  - \frac{7}{8m^4c^4 Z \left(1 + Q/mc^2 \right)^8} \sum_{j = 1}^Z \langle \widetilde 0 \vert p_j^4 \vert \widetilde 0 \rangle,
\end{eqnarray}

\noindent and

\begin{eqnarray}
	\label{deltaV}
	\Delta_{_V}(Q)  =  -\frac{\hbar (1 + Q/3mc^2)}{2m^3c^4Z(1 + Q/mc^2)^4(1+Q/2mc^2)} \sum_{j = 1}^Z \langle \widetilde 0 |\vec \sigma_j \cdot ( \nabla_j V_j \times \vec p_j)| \widetilde 0 \rangle,
\end{eqnarray}

\noindent Note that in $\Delta_{_V}(Q)$, the coefficient of $\nabla_j^2 V_j$ vanishes identically. This is somewhat surprising, given the complicated nature of the calculations leading to this result. Expectation values of the remaining spin-orbit operator are usually  smaller then some of the two-body terms that arise from $V_{kl}$, and which I have omitted \cite{stop_one,myRange} because they are known to be quite small \cite{FragaSO,FragaOO,FragaMV}. Therefore, to a high level of accuracy we have

\begin{equation}
	\label{deltaV_0}
	\Delta_{_V}(Q) = 0,
\end{equation}

\noindent and the potential energy of the electrons does not appear explicitly in the result, entering only through its effects on the ground-state wavefunction of the target system. These results give the relativistic corrections to ${\cal O}(p^4/m^4c^4)$ and ${\cal O}(p^2V/m^3c^4)$. Since they are valid to all orders in $Q/mc^2$, they include contributions from all multipoles, unlike earlier studies \cite{Leung+,mysumrule}. It should also be remembered that the Bethe sum arises from the longitudinal part of the DCS and as such, does not involve retardation, in contrast to the photoabsorption sum of GGT.

The TRK sum rule may be obtained directly from the above results as $S_1 = Z(1 - \Delta_f)$, with \cite{myRange}

\begin{eqnarray}
	\label{delta}
	\Delta_f = \frac{5}{6m^2c^2Z} \sum_{j = 1}^Z \langle \widetilde 0|p_j^2| \widetilde 0 \rangle  - \frac{7}{8m^4c^4 Z} \sum_{j = 1}^Z \langle \widetilde 0 \vert p_j^4 \vert \widetilde 0 \rangle.
\end{eqnarray}

\noindent The first term is just that found by numerous others \cite{Levinger,Dogliani,Matsuura,Friar,Schmitt}, but the second term had not previously been known. The reader may recall that the TRK sum rule for a single electron in the field of a $Pb$ nucleus was evaluated (along with electric quadrupole and retardation contributions) using perturbative methods in \cite{Levinger}. It is consistent with these methods to evaluate the $p^4$ term using an NR wavefunction. For the one-electron case and a $Pb$ nucleus, this gives a contribution of $-35 (Z\alpha)^4/8 \approx -0.53$, whereas the first-order term is $0.30$. Thus, it is quite clear that this approach fails badly for such strongly bound systems. For neutral atoms, on the other hand, we may expect that the average over electrons will diminish the errors and allow the perturbative approach to give a reasonably close approximation. This notion is supported by the results I obtained in \cite{myRange}. Nonetheless, it needs to be checked with the more accurate approach described below, using DKT wavefunctions.

Numerical studies of Eqs.~(\ref{deltaf}) and (\ref{delta}) indicate that perturbative calculations of these sum rules for neutral atomic systems yield accurate results over a wide range of $Z$ \cite{myRange}. Further work is needed, however, when $Z > 70$ and $Q$ is small, if one wishes to achieve an accuracy of $0.5\%$ or better \cite{myRange}. To obtain this greater accuracy for high $Z$, it is possible to use the methods already described, but to forego completely the expansion in powers of $p_j$. That is, we may insert Eqs.~(\ref{FWfree}), (\ref{H}), (\ref{Hq}), and (\ref{Ujq}), directly into Eq.~(\ref{myBethesum}), and evaluate the expectation value using the accurate DKT ground state of $\widetilde H$. Expectation values using these states, and carefully including the effects of $V_{kl}$, have been used extensively by Hess and co-workers \cite{Hess1,Hess3,Hess4}. The success these authors have had with such calculations, keeping only low-order terms in $V$, makes it quite likely that this approach will also yield highly accurate results for the sum rules.

For the ${\cal O}(V)$ terms calculated without the $p$-expansion, numerical evaluation of the Bethe sum for nonzero $Q$ appears to offer quite a significant challenge \cite{HessComm}. Therefore, I have begun with consideration of the TRK sum, $S_1 = Z(1 - \Delta_f - \Delta_V)$, keeping only ${\cal O}(V)$ in $\Delta_V$. Now, $\Delta_f$ has the relatively simple form,

\begin{equation}
	\label{exactTRK}
	\Delta_f = \frac{mc^2}{3Z}\sum_{k=1}^Z \left\langle \widetilde 0\left|\left(\frac{2}{\epsilon_k} +\frac{m^2c^4}{\epsilon_k^3}\right) \right| \widetilde 0 \right\rangle ,
\end{equation}

\noindent but the expression for $\Delta_V$ is quite complicated, so will not be written out here. Though accurate numerical calculations of $\Delta_f$ and $\Delta_V$ are yet to be completed, I have obtained order-of-magnitude estimates of these quantities for a single-electron system by using NR hydrogenic wavefunctions. The results are intriguing, and they will be the basis of the discussion in the remainder of this section, which will mainly be speculative in nature pending the outcome of work in progress.

For all values of $Z$, these rough estimates indicate that $\Delta_V$ is much smaller than $\Delta_f$. Considering once again a single electron in the field of a $Pb$ nucleus, I find $\Delta_f = 0.17$, while $\Delta_V = 0.002$ . If instead we have a nucleus with $Z = 120$, then $\Delta_f = 0.27$ and $\Delta_V = 0.008$. We see that even for such extreme conditions, $\Delta_V / \Delta_f  \approx 0.03$, and $\Delta_V$ itself is only about $1\%$ of the total sum. For most physical systems, only a small fraction of the electrons will have binding energies approaching those just considered. For neutral atoms, I have found $\Delta_f$ to be no more than about $0.03$ in the perturbative approach \cite{stop_one,myRange}. If  the relative size of the two terms remains even approximately the same, than we will be justified in neglecting $\Delta_V$. If this is confirmed by the more accurate numerical methods using many-electron DKT wavefunctions, then Eq.~(\ref{exactTRK}) will give us the effectively exact relativistic corrections, $E_1^{rel}$, to the TRK sum rule. Since this expression does not involve the potential energy $V$ and is rather simple in form, it should be relatively easy to evaluate even with realistic wavefunctions.

We may hope for more. In studies of the Bethe sum, it was found that the relativistic corrections were never much larger than their value at $Q = 0$ \cite{myRange}. Therefore, given the relationship between the TRK sum and the $Q \rightarrow 0$ limit of the Bethe sum (see Eq.~(\ref{S1})), it seems probable that $\Delta_V(Q)$ will be negligible over the whole range of $Q$. If this can be demonstrated, then the corrections to the Bethe sum rule will be given to high accuracy by setting $V = 0$ in the operators appearing in the expectation value of Eq.~(\ref{myBethesum}). This reduces to the simple expression \cite{stop_one,mysumrule},

\begin{equation}
	S_1(Q) = \frac{c^2}{2}\sum_{j=1}^Z \left\langle \widetilde 0 \left\vert \left( \frac{q^2 + \vec q \cdot \vec p_j}{\epsilon_j(\vec q)} + \frac{\vec q \cdot \vec p_j}{\epsilon_j}\right) \right\vert \widetilde 0 \right\rangle.
\end{equation}

\noindent Then this approach, using the DKT without expanding in powers of $p$, could be used to obtain accurate stopping powers for targets consisting of even the heaviest elements.

\subsection{Photoabsorption sum rules for scattering of charged particles}
	\label{trans}

In this section, I will discuss the summations that arise from considerations of the transverse part of the DCS, given in Eq.~(\ref{sigma}), and show they are closely related to the GGT sum. In \cite{stop_one}, I have evaluated sum rules for these contributions to the stopping power using both low- and high-$Q$ approximations. In the high-$Q$ region, the recoil energy and energy transfer cannot differ significantly \cite{InokutiRMP,Fano}, so it is a good approximation to make an expansion in powers of $(Q - E_{ni})/Q$, which leads to sums of the form given in Eq.~(\ref{tsums}). Although these sums involve the Dirac operators $\vec \alpha_j$, they can be handled with the same techniques described in the previous section for treatment of the longitudinal contributions. Details of this approach, along with corresponding results  for the stopping power, are given in \cite{stop_one}.

The low-$Q$ transverse contributions to the stopping power are given by

\begin{equation}
	\label{lowq}
	B_t = \frac{1}{2} \sum_{n} E_{ni} \int_{Q_{m}}^{Q_{c}} \left[ \frac{|\vec \beta_t \cdot\vec G_n|^2}
			{[Q \left(1+Q/2mc^2\right)- E_{ni}^2/2mc^2]^2} \right] \left(1 + Q/mc^2 \right) dQ.
\end{equation}

\noindent  Fano's treatment of $B_t$ began with the change of variables \cite{Fano},

\begin{equation}
	\label{psi}
	\cos^2 \psi = \frac{Q_{m} (Q_{m} + 2mc^2)}{Q (Q + 2mc^2)},
\end{equation}

\noindent giving $q = E_{ni}/v\cos \psi$ and

\begin{equation}
	\label{LowL}
	B_t = \sum_{n} \frac{mc^2 \beta^4}{Z E_{ni}} \int_{\cos^2 \psi_c}^{1} \frac{\sin^2 \psi}{[1 - \beta^2 \cos^2 \psi]^2} \left| \sum_{j=1}^Z \langle n | e^{iq z_j / \hbar} \alpha_{jx} | i \rangle \right |^2 d(\cos^2 \psi).
\end{equation}

\noindent Here, $\psi$ is the angle between $\vec \beta$ and $\vec q$ so that $\beta_t = \beta \sin \psi$, and $\psi_c$ is the value of $\psi$ when $Q = Q_c$. Fano then set $q = 0$ in the exponential and related the matrix element of $\alpha_x$ to that of $x$, reducing this expression to one proportional to the dipole oscillator strength, $f_n$ \cite{Fano}.

In obtaining the relativistic corrections to the stopping power, I found it necessary to keep terms ${\cal O} (q^2)$ from the exponential in Eq.~(\ref{LowL}) in order that the results for the stopping power were independent of the choice of cutoff, $Q_c$ \cite{stop_one}. Rewriting $q$ in terms of $E_{ni}$ and expanding the exponential allows the $\psi$-integral to be done, and yields an expression for $B_t$ involving the two sums,

\begin{equation}
	\label{fnrel}
	\sum_n \frac{1}{E_{ni}} \left| \sum_{j=1}^Z \langle n | \alpha_{jx} | i \rangle \right |^2 
\end{equation}

\noindent and

\begin{eqnarray}
	\label{Gn3}
	\sum_n E_{ni} \left[ \left| \sum_{j=1}^Z \langle n |  z_j \alpha_{jx} | i \rangle \right |^2 - \frac{1}{2} \left( \sum_{j,k=1}^Z \langle i | \alpha_{kx} | n \rangle \langle n | z_j^2 \alpha_{jx} | i \rangle + H.c. \right)\right],
\end{eqnarray}

\noindent where the terms linear in $z_j$ vanish due to conservation of parity. Since $[H,x_j] = -i\hbar c\alpha_{jx}$ (apart from small two-body terms), Eq.~(\ref{fnrel}) is just the relativistic TRK sum, representing the unretarded dipole contributions $E1$. Comparison to Eq.~(\ref{pz}) reveals that Eq.~(\ref{Gn3}) is just the relativistic form of the combination, $E2 + E1^{ret}$, considered in the work on photoabsorption sum rules. It is thus seen to have been no accident that for a relativistic treatment of the target, I needed to keep higher-order terms in the expansion of the exponential of Eq.~(\ref{LowL}). As has been pointed out previously, the effects of higher multipoles, retardation, and relativity all lead to contributions of the same order of magnitude. Keeping only the leading term in the expansion excludes the $E2$ and $E1^{ret}$ contributions, and cannot yield a consistent relativistic formulation. We also see here a direct relationship between the GGT sum, which arose out of studies of photoabsorption, and the sums related to the transverse contributions to the stopping power. 

Though we have arrived at this conclusion from consideration of the lowest order multipoles, it can be shown that this relationship holds for all multipoles. If the lower limit of the integral in Eq.~(\ref{LowL}) is avoided, then we may directly extract from Eq.~(\ref{LowL}) the sum,

\begin{equation}
	\label{transGGT}
	\sum_{n} \frac{1}{E_{ni}} \left| \sum_{j=1}^Z \langle n | e^{iqz_j/\hbar} \alpha_{jx} | 0 \rangle \right |^2,
\end{equation}

\noindent which is just the relativistic generalization ($\vec p \rightarrow \vec \alpha$) of the integrated photoabsorption cross-section in Eq.~(\ref{abs}). Thus, the sums for the transverse contributions to the stopping power have contributions that are exactly the same as those appearing in the photoabsorption dispersion relation of GGT.

We may understand the appearance of the photoabsorption sum in the context of charged particle scattering in the following way:  First, note that this contribution to the stopping power represents the exchange of transverse photons between projectile and target. If the target is initially in its ground state, it cannot emit, and can only absorb. Then, observe that in Eq.~(\ref{LowL}), the projectile states have already been integrated out, leaving a description in terms of the target states alone. Thus, this equation describes the absorption of photons by the target, and it is completely reasonable that it contains the photoabsorption sum.

Nonetheless, the reader may note two important differences between Eqs. (\ref{abs}) and (\ref{LowL}) . First, for the charged particle case, the energy-momentum relation given after Eq.~(\ref{psi}) is not that usually seen for photons. This is because these are virtual photons, for which energy and momentum can take on any value, and each may do so independently of the other. One may imagine that both projectile and target are constantly emitting and reabsorbing virtual photons, and occasionally one of these photons gets exchanged between the two. The photons that are exchanged must be ones that lead to conservation of both energy and momentum in the overall interaction between projectile and target, since that is a real process. This requirement leads to the relation, $q = E_{ni} / v \cos \psi$ (except at extremely high incident energies, approaching the $TeV$ range for protons) instead of $E_{ni}/c$.

The second difference has to do with the presence of the $\psi$-integration in Eq.~({\ref{LowL}). This integral simply reflects the fact that contributions from all possible final states of the projectile have already been summed. Therefore, unlike in the photoabsorption case where there is only one possible momentum transfer for each $E_{ni}$, here there is a range of momentum transfer, or equivalently $\psi$, that is possible. One may also note that the upper limit, $\cos^2 \psi_c$, of the $\psi$-integral depends on $E_{ni}$. Therefore, there are additional terms that differ in form from those resembling the photoabsorption sum.

\section{The trouble with relativity}
\label{relativity}

We have seen that significant challenges arise in the evaluation of sum rules for relativistic systems. Unlike their non-relativistic counterparts, the use of closure for relativistic sum rules requires the introduction of special techniques, such as the FWT or projection operators \cite{Schmitt}, to exclude contributions from negative energy states. Direct use of closure without such techniques in a single-particle theory of a hydrogenic system includes contributions from transitions of the electron into negative energy states. As previously stated, these contributions precisely cancel those from transitions into positive energy states \cite{Levinger}, yielding unphysical vanishing results for the sum rules.

If one views these sums from the many-electron perspective of Dirac's hole theory, on the other hand, the negative energy states are filled (Dirac sea) and transitions into these states are automatically excluded. Then one might expect an absence of such difficulties when using the closure relation, $\sum_n |n \rangle \langle n | = {\cal I}$, with $|n\rangle$ now a many-electron state including the filled Dirac sea. However, one still obtains a vanishing result when an average over the direction of $\vec q$ is taken (the mathematics is essentially the same as for single-electron theory). The explanation of this result is no longer straightforward, but as I will now describe, it is related to the appearance of infinities from the participation of electrons in the Dirac sea.  For the following discussion let us consider the Bethe sum, 

\begin{equation}
	S_{1}(Q) = \sum_{j} \sum_{n} E_{ni} \left| \langle n \vert e^{i \vec{q} \cdot \vec r_j / \hbar} \vert i \rangle \right|^2,
 	\label{S1Qrepeat}
\end{equation}

\noindent and now $\sum_j$ includes all negative energy electrons, as well as those with positive energy. For simplicity, I will restrict the discussion to a single, free electron at rest in the presence of the Dirac sea. Then with $|i\rangle$ and $|n\rangle$ given by suitably symmetrized sets of momentum states, the matrix elements for a given $\vec q$ are proportional to $\delta(\vec p_j \,^\prime - \vec p_j - \vec q)$, and only one final state contributes to the sum over states for each $j$. For bound electrons, this probability distribution will be broadened, as has been discussed in reference to the Bethe surface \cite{InokutiRMP,Fano,RauFano}, but the basic arguments presented below still hold. Therefore, the conclusions should be quite general.

In consideration of this system, we may identify two possible sources of infinity. The first is the infinite energy, $E_D$, of the Dirac sea contained in each of $E_n$ and $E_i$. The other is the fact that $\sum_j$ runs over an infinite number of electrons. Let us consider each of these in turn.

For a given $j$, we have a sum over states $|n \rangle$ and a subtraction of the energies, $E_{ni} = E_n - E_i$. If the subtraction is taken first, explicitly calculating each individual term prior to the summation over states, there is a precise cancelation of the infinite $E_D$ between $E_n$ and $E_i$. Then a finite result is obtained, which is proportional to $q^2$ for small $q$ (this calculation is straightforward for the free-particle states presently under consideration). 

On the other hand, when the sum over states is done first using closure, we obtain a very different result,

\begin{eqnarray}
	S_1(Q)  =  \sum_j  \langle i \vert e^{-i \vec{q} \cdot \vec r_j / \hbar}  [H, e^{i \vec{q} \cdot \vec r_j / \hbar}] \vert i \rangle  =  c \sum_j  \langle i \vert \vec \alpha_j \cdot \vec q \vert i \rangle.
\end{eqnarray}

\noindent For each $j$, the result is again finite, but here it is linear in $\vec q$ and therefore, upon averaging over the direction of $\vec q$, we obtain the vanishing result referred to above. What has happened is that we have reformulated the sum into a difference of two terms, 

\begin{eqnarray}
	\sum_n E_n |\langle n | e^{i \vec{q} \cdot \vec r_j / \hbar} | i \rangle|^2 - \sum_n E_i |\langle n | e^{i \vec{q} \cdot \vec r_j / \hbar} | i \rangle|^2 = \langle i \vert H(\vec{q}) \vert i \rangle - \langle i \vert H \vert i \rangle,
\end{eqnarray}

\noindent each of which is infinite. While the infinities still cancel, the resulting finite result is of a very different form as that found by explicit calculation of each individual term, followed by the summation. Because of the presence of infinities, it is incorrect to interchange the order of these mathematical operations, the subtraction and summation being considered here.

One may ask, which method is the correct one? We will argue that it is the first one, where the subtraction is taken before the summation, even though, as we shall see, this yields an infinite result. 

To begin with, I remind the reader that the Bethe sum arose in a calculation of the mean energy loss from fast charged particles to matter. The definition of this sum thus involves terms, each of which has a factor of the energy transfer, $E_{ni}$. Each term to be summed is proportional to this energy difference; that is, according to the original derivation of the sum, the difference should be taken before the summation over states. As further evidence, note that our result should reduce to the NR result in the appropriate limit. The NR result is also proportional to $q^2$ and does not vanish upon averaging over the direction of $\vec q$. Therefore, we conclude that the correct result is that obtained when the summation over states comes after the subtraction of energies. The alternative, where closure is used directly to obtain a sum rule, is seen to be faulty, as a result of the interchange of mathematical operations in the presence of infinities.

Next let us consider the sum over electrons. As I have said, for any given free electron only a single transition contributes to the sum, that for which $\vec p_j\,^\prime = \vec p_j + \vec q$. For a positive energy electron initially at rest, the energy difference between these states is just $Q \le Q_M$. Transitions involving energies greater than $Q_M$ cannot occur. In this sense, though energy conservation has been discarded by including all eigenstates of the Hamiltonian in the summation, it has been effectively reinstated by the upper limit of the $Q$-integration.

With the Dirac sea of electrons, we are not so lucky. For each negative energy state of momentum $\vec p_j$, there is a corresponding positive energy state of momentum $\vec p_j + \vec q$, and the energy difference between these states can be as large as $\,2|E_i|$ or larger. Since summation over the entire Dirac sea includes the limit $|E_i| \rightarrow \infty$, transitions involving the production of arbitrarily high energy electron-positron pairs make significant contributions to the sum. Although the corresponding matrix elements decrease with increasing $|E_i|$, they do not decrease fast enough to keep the sum over electrons from diverging. An explicit calculation, readily done for the free-electron model, shows that the resulting sum is infinite.

Though these arguments have specifically considered a free-electron model, the basic ideas hold for bound electrons, as well. The transition probability will in this case be broadened by the binding energy, but pairs of arbitrarily high energy will still be produced for any given recoil energy $Q$. Therefore, the conclusions of the previous paragraphs should be valid generally. The same conclusions also apply to the TRK sum, since it may be obtained directly from the Bethe sum, as shown in Eq.~(\ref{S1}). In the photoabsorption sum of GGT the momentum transfer is fixed by the relationship, $q = E_{ni}/c$. While this differs from that in the free-electron model for charged particle scattering, the above arguments do not depend on the precise relationship between these quantities. Therefore, the arguments are applicable to all sum rules considered in this article.

I wish to stress that it is completely \lq \lq reasonable" that the result is infinite. The use of closure has removed any limits on the energy transfer, and the model no longer obeys conservation of energy, at least not with respect to the finite energy of any realistic physical system. This point is of some interest, since it seems to imply that the presence of infinities is an unavoidable consequence of the use of closure along with inclusion of pair production processes in these sums. The sums include contributions from creation of arbitrarily high energy electron-positron pairs. Furthermore, these are real, as opposed to virtual,  pairs. It is not, for example, a matter of an electron being surrounded by a sea of virtual pairs, leading to polarization of the vacuum and altering observable quantities, such as the electron's charge and mass. Rather, these are real  pairs being produced by collision with a projectile, which in this treatment is effectively being considered to have infinite energy. Since there is infinite energy available, real pairs of unlimited energy can be produced, and the infinite result for the mean energy transfer is in this sense quite reasonable. Though one might have hoped that an approach utilizing the full power of quantum electrodynamics, including the removal of infinities by the methods of renormalization, would allow the effects of pair production to be incorporated along with the usual excitation of positive energy electrons, these observations seem to indicate that any such efforts will prove unsuccessful if they also utilize closure to obtain sum rules. It appears that calculations of the pair production contributions to these sums will require an approach with explicit limits on the amount of energy exchanged between projectile and target.

\section{Conclusion}
\label{remarks}

In this article, I have discussed relativistic effects on sum rules appearing in the description of scattering of charged particles and photons from atoms. Having begun with considerations of how sum rules arise in these contexts, I then reviewed the early work on this challenging subject. Following this, I discussed my own recent contributions utilizing the Foldy-Wouthuysen and Douglas-Kroll transformations, and conjectured the latter has the capability of producing relatively simple expressions that give highly accurate results for these sums for even the heaviest of elements. I also demonstrated the equivalence of sums for the transverse interaction in the charged particle case and those which describe photoabsorption. Finally, I have discussed the difficulties that arise in evaluating these sum rules from the perspective of Dirac's hole theory, and shown that the understanding of these difficulties is much less straightforward then when considered from a single-particle point of view. In hole theory, the difficulties were traced to infinities that arise due to the presence of the filled Dirac sea. These considerations suggest that it will not be possible to obtain physically reasonable sum rules using closure if pair production processes are also included in the treatment.

\section*{Acknowledgments}
\label{sec:ack}

I would like to offer my thanks to the members of the Physics Department at Lewis and Clark College for their gracious hospitality during my summer stay in Portland where this paper was written. Acknowledgment is also made to the donors of The Petroleum Research Fund, administered by the American Chemical Society, for support of this research.


\newpage

\begin{thebibliography}{10}

\bibitem{InokutiRMP}
M. Inokuti, Rev. Mod. Phys. {\bf 43},  297  (1971).

\bibitem{Dalgarno}
A. Dalgarno, Rev. Mod. Phys. {\bf 35},  522  (1963).

\bibitem{Jackiw}
R. Jackiw, Phys. Rev. {\bf 157},  1220  (1967).

\bibitem{Levinger}
J.~S. Levinger, M.~L. Rustgi, and K. Okamoto, Phys. Rev. {\bf 106},  1191
  (1957).

\bibitem{Fano}
U. Fano, Ann. Rev. Nucl. Sci. {\bf 13},  1  (1963).

\bibitem{Bethe1}
H. Bethe, Ann. Physik {\bf 5},  325  (1930).

\bibitem{Bethe3}
H. Bethe,  in {\em Handbuch der Physik} (Springer, Berlin, 1933), Vol.~24/1,
  p.\ 273.

\bibitem{Walske1}
M. Walske, Phys.Rev. {\bf 88},  1283  (1952).

\bibitem{Walske2}
M. Walske, Phys.Rev. {\bf 101},  940  (1956).

\bibitem{BichselShell}
H. Bichsel, Phys. Rev. A {\bf 65},  052709  (2002).

\bibitem{BetheRMP}
M. Livingston and H. Bethe, Rev. Mod. Phys. {\bf 9},  245  (1937).

\bibitem{BetheJackiw}
H. Bethe and R. Jackiw, {\em Intermediate Quantum Mechanics}, 2nd ed.
  (Benjamin, New York, 1968), Chap.~11.

\bibitem{stop_one}
S.~M. Cohen, Phys. Rev. A {\bf 68},  012720  (2003).

\bibitem{GGT}
M. Gell-Mann, M.~L. Goldberger, and W.~E. Thirring, Physical Review {\bf 95},
  1612  (1954).

\bibitem{BrownPrComm}
G. E. Brown, private communication referenced in \cite{LevPayne,LevRustgi}.

\bibitem{Erber}
T. Erber, Annals of Physics {\bf 6},  319  (1959).

\bibitem{LevPayne}
W.~B. Payne and J.~S. Levinger, Phys. Rev. {\bf 101},  1021  (1956).

\bibitem{LevRustgi}
J.~S. Levinger and M.~L. Rustgi, Phys. Rev. {\bf 103},  439  (1956).

\bibitem{Brown1}
S. Brenner, G.~E. Brown, and J.~B. Woodward, Proc. Roy. Soc. London Ser. A {\bf
  227},  59  (1954).

\bibitem{Brown2}
G.~E. Brown and D.~F. Mayers, Proc. Roy. Soc. London Ser. A {\bf 234},  387
  (1956).

\bibitem{Hulme}
Hulme, McDougall, Buckingham, and Fowler, Proc. Roy. Soc. London Ser. A {\bf
  149},  131  (1935).

\bibitem{Jacobsohn}
B. Jacobsohn, Ph.D. dissertation, University of Chicago, 1947 (unpublished).

\bibitem{Pratt1}
R.~H. Pratt and Y.~S. Kim, Romanian Journal of Physics {\bf 38},  353  (1993).

\bibitem{Pratt2}
A. Ron {\it et~al.}, Phys. Rev. A {\bf 50},  1312  (1994).

\bibitem{GoldLow}
M.~L. Goldberger and F.~E. Low, Physical Review {\bf 176},  1778  (1968).

\bibitem{Dogliani}
H.~O. Dogliani and W.~F. Bailey, J. Quant. Spectrosc. Radiat. Transfer {\bf 9},
   1643  (1969).

\bibitem{Friar}
J.~L. Friar and S. Fallieros, Phys. Rev. C {\bf 11},  274  (1975).

\bibitem{Matsuura}
T. Matsuura and K. Yazaki, Phys. Lett. {\bf 46B},  17  (1973).

\bibitem{Gerasimov}
S.~B. Gerasimov, Phys. Lett. {\bf 13},  240  (1964).

\bibitem{Drake}
S. Goldman and G. Drake, Phys. Rev. A {\bf 25},  2877  (1982).

\bibitem{Foldy}
L.~L. Foldy and S.~A. Wouthuysen, Phys. Rev. {\bf 78},  29  (1950).

\bibitem{Schmitt}
K.-M. Schmitt and H. Arenh\"ovel, Z. Phys. A {\bf 320},  311  (1985).

\bibitem{Leung+}
P.~T. Leung, M.~L. Rustgi, and S.~A.~T. Long, Phys. Rev. A {\bf 33},  2827
  (1986).

\bibitem{Leung}
P.~T. Leung, Phys. Rev. A {\bf 40},  5417  (1989).

\bibitem{mysumrule}
S.~M. Cohen and P.~T. Leung, Phys. Rev. A {\bf 57},  4994  (1998).

\bibitem{Sucher}
J. Sucher, Physical Review A {\bf 22},  348  (1980).

\bibitem{Sabin}
G.~A. Aucar, J. Oddershede, and J.~R. Sabin, Phys. Rev. A {\bf 52},  1054
  (1995).

\bibitem{DouglasKroll}
M. Douglas and N. Kroll, Ann. Phys. (N.Y.) {\bf 82},  89  (1974), see pp.
  110-113 and Appendix III beginning on p. 151.

\bibitem{Hess1}
B. Hess, Phys. Rev. A {\bf 33},  3742  (1986).

\bibitem{Hess2}
G. Jansen and B. Hess, Phys. Rev. A {\bf 39},  6016  (1989).

\bibitem{myRange}
S.~M. Cohen, scheduled to appear in Phys. Rev. A, October, 2003.

\bibitem{FragaSO}
B. Lo, K. Saxena, and S. Fraga, Theoret. chim. Acta (Berl.) {\bf 25},  97
  (1972).

\bibitem{FragaOO}
K. Saxena, B. Lo, and S. Fraga, J. Phys. B {\bf 5},  768  (1972).

\bibitem{FragaMV}
B. Lo, K. Saxena, and S. Fraga, Theoret. chim. Acta (Berl.) {\bf 24},  300
  (1972).

\bibitem{Hess3}
A. Wolf, M. Reiher, and B.~A. Hess, J. Chem. Phys. {\bf 117},  9215  (2002).

\bibitem{Hess4}
A. Wolf, M. Reiher, and B.~A. Hess,  in {\em Relativistic Quantum Chemistry},
  edited by P. Schwerdtfeger (Elsevier Science, Amsterdam, 2002), Vol.~24/1,
  pp.\ 622--663.

\bibitem{HessComm}
B. A. Hess, private communication.

\bibitem{RauFano}
A. Rau and U. Fano, Phys. Rev. {\bf 162},  68  (1967).

\end{thebibliography}
\end{document}